\documentclass[aps,prb,showpacs,groupedaddress]{revtex4}
\usepackage{amsmath,graphicx}
\usepackage{epsfig}
\usepackage[next]{inputenc}

\begin{document}
\title{Green's function approach to quantum criticality\\ in the anisotropic Kondo necklace model}

\author{H. Rezania}
\affiliation{Physics Department, Sharif University of Technology, Tehran 11155-9161, Iran}
\author{A. Langari}
\email[]{langari@sharif.edu}
\homepage[]{http://spin.cscm.ir}
\affiliation{Physics Department, Sharif University of Technology, Tehran 11155-9161, Iran}
\author{P. Thalmeier}
\affiliation{Max Planck Institute for Chemical Physics of Solids, 01187 Dresden, Germany}
\begin{abstract}
We have studied the quantum phase transition between the antiferromagnetic and 
spin liquid phase  for the two dimensional anisotropic Kondo
necklace model. The bond operator formalism has been implemented to transform the spin
Hamiltonian to a bosonic one. We have used the Green's function approach including a hard
core repulsion to find the low energy excitation spectrum of the model. The bosonic excitations
become gapless at the quantum critical point where
the phase transition from the Kondo singlet state to long range antiferromagnetic order takes place.
We have studied the effect of both inter-site ($\delta$) and local ($\Delta$) anisotropies
 on the critical point and on the critical exponent of the excitation gap in the paramagnetic phase.
We have also compared our results with previous bond operator mean field calculations.
\end{abstract}
\date{\today}
\maketitle

\section{Introduction}
 The description of quantum phase transition between phases with spontaneously
 broken symmetry and disordered phases is a novel topic in condensed
 matter physics\cite{sachdev,vojta}. Macroscopic strongly correlated electron systems at low temperature  (and as a function of magnetic field, hydrostatic or chemical pressure) show a wide range of interesting phenomena, such as quantum criticality and associated  non-Fermi liquid (NFL) behaviour, magnetism, Kondo insulating behaviour and  superconductivity \cite{hewson,stewart}. In the single-impurity case the Kondo problem describes the antiferromagnetic interaction ($J$) between the impurity spin and the free conduction electron spins. This gives rise to a new non-perturbative low-energy scale,  the Kondo temperature  $T_{K}$ which dominates the low temperature anomalies in the thermodynamic and transport quantities\cite{hewson}. $T_K=De^{-1/(2J\rho)}$ (D, $\rho$ are conduction band width and density of states, respectively) has the meaning of a crossover temperature from uncoupled local spins for T $\gg$ T$_K$ to the strongly coupled local spins, forming a singlet ground state with conduction electrons, for T $\ll$ T$_K$. In the Kondo lattice (KL) model  \cite{tsunetsugu} an additional (perturbative) energy scale $T_{RKKY}=J^{2}\rho$ for the effective Ruderman-Kittel-Kasuya-Yosida (RKKY) 
inter-site interactions of local spins appears. This model exhibits a quantum phase transition  between the Kondo singlet phase and the magnetically ordered phase as function of the control parameter $x=J\rho(E_F)$ as argued by Doniach\cite{doniach}. The transition takes place at a quantum critical point (QCP) characterised by $x_c=J_c\rho(E_F)$ where $x_c$ is of the order one. For $x\ll x_c$ the effective interactions dominate and magnetic order appears. For $x\gg x_c$ the singlet formation dominates and a heavy Fermi liquid state is realized. This qualitative picture has been supported by numerical calculations within dynamical mean field theory (DMFT) and exact diagonalization methods \cite{sun,zerec} for the Anderson lattice and Kondo lattice Hamiltonian respectively. However the vicinity of the quantum critical point and associated NFL behavior \cite{schroeder} requires a treatment within phenomenological effective models as developed in Refs.~\onlinecite{hertz,millis}.  
 
The Kondo lattice model emerges from periodic Anderson model via a Schrieffer-Wolff transformation that eliminates
the fluctuations of f-charge or f-orbital occupation  \cite{tsunetsugu} . It is given by
\begin{equation}
 H_{KL}=t\sum_{\left\langle ij\right\rangle  ,\tau}(c^{\dagger}_{i,\tau}c_{j,\tau}+h.c.)
 +J_{\bot}\sum_{i}{\bf \tau}_{i}{\bf S}_{i} \;\;. 
\label{e2}
\end{equation}
The first part describes conduction electrons $c^{\dagger}_{i,\tau}$ with n.n. hopping t. The second part is the Kondo term
where ${\bf \tau}_{i}$ and ${\bf S}_{i}$ are conduction electron and localized spin respectively. This model still contains the charge fluctuations of conduction electrons  expressed by the hopping term. It was shown by Doniach \cite{doniach} that in 1D it may be replaced by an xy-type inter-site exchange term. Thus the Kondo lattice model is replaced by a pure spin Hamiltonian, the Kondo-necklace model (KNM). In higher dimension this procedure cannot be justified strictly. However suppose we add a Coulomb repulsion U$_c$ between conduction electrons to the KL model (Eq.~\ref{e2}) in the half filled case. Then in the limit $U_c/t\rightarrow\infty$ charge fluctuations of conduction electrons are frozen out and the low energy physics is again described by a pure spin Hamiltonian. Strictly speaking this is only adequate for the Kondo insulator with a charge gap but one may expect that it is also useful to describe the low energy spin dynamics of metallic Kondo systems. The generalized Kondo necklace model obtained in this way \cite{bruenger} is given by
\begin{equation}
H=J\sum_{{\langle i,j\rangle}}
(\tau^{x}_{i}\tau^{x}_{j}+\tau^{y}_{i}\tau^{y}_{j}+\delta \tau^{z}_{i}\tau^{z}_{j})
+J_{\perp}\sum_{\langle
i \rangle}(\tau^{x}_{i}S^{x}_{i}+\tau^{y}_{i}S^{y}_{i}+\Delta \tau^{z}_{i}S^{z}_{i}) \;\;,
\hspace{5mm}
\label{e10}
\end{equation}
where the intersite exchange $J$ is of order  $t^2/U_c$. In 2D which will be considered in the present work this is equivalent to a special case of an (anisotropic) bilayer-Heisenberg model \cite{kotov} where the inter-site bonds (J) are cut in one layer and J$_\perp$ is the inter-layer coupling. Here  both  spins are 1/2 and the exchange coupling parameters are
antiferromagnetic $(J,J_{\perp}\geq 0)$. 
In the above Hamiltonian, $\tau_{i}^{\alpha}$ represent
the $\alpha$-component of spin of the 'itinerant' electrons at 
site $i$ and $S_{i}^{\alpha}$ is 
the $\alpha$-component of localized spins at position $i$. 

We want to study the possible quantum phase transition of this model under rather general assumption of both anisotropies in the inter-site interaction ($\sim J$) and on-site Kondo terms ($\sim J_\perp$) of Eq.~(\ref{e10}). They are characterized by a pair of parameters ($\delta,\Delta$). The $\Delta$ anisotropy is always present in real Kondo compounds like Ce-based intermetallics due to the crystalline electric field (CEF) and $\delta$ is caused by spin-orbit coupling of conduction electrons. We study the quantum phase transition from the paramagnetic (Kondo-singlet) side as function of the control parameter $J_{\bot}/J$, which gives the ratio of the inter-site to the on-site interaction strength, and as function of the anisotropy parameters ($\delta,\Delta$). 
We have implemented the Green's function approach introduced to study 
the bilayer isotropic Heisenberg model ($\delta,\Delta$) = ($1,1$) \cite{kotov}. 
The effect of anisotropies on the quantum phase transition of KNM has recently been studied
by a mean field approach both in the absence \cite{langari} and presence \cite{thalmeier} 
of a magnetic field. However, using the more advanced Green's function method we will obtain more accurate values for the critical gap exponents $\nu$ and  for quantum critical point values  $(J_\perp/J)_c$ which differ both from the mean field values.



\section{Boson operator representation of the model Hamiltonian}

\label{sect:model}

The bond operator representation  introduced by Chubukov\cite{chubukov} and Sachdev, et.al \cite{bhatt}  is a useful approach to describe disordered phases. This representation can be considered as an
analog of the usual Holstein-Primakov transformation for phases with broken spin rotational symmetry. In terms of singlet-triplet operators, the spin operators of the localized and conduction electrons are given by 
\begin{eqnarray}
&&S_{i,\alpha}=\frac{1}{2}(s^{\dagger}_{i}t_{i,\alpha}+t_{i,\alpha}^{\dagger}s_{i}-i\epsilon_{\alpha\beta\gamma}t_{i,\beta}^{\dagger}t_{i,\gamma}) \;,\nonumber \\
&&\tau_{i,\alpha}=\frac{1}{2}(-s^{\dagger}_{i}t_{i,\alpha}-t_{i,\alpha}^{\dagger}s_{i}-i\epsilon_{\alpha\beta\gamma}t_{i,\beta}^{\dagger}t_{i,\gamma})\;\;,
\label{e5}
\end{eqnarray}
where $(\alpha,\beta,\gamma)$ represent the (x,y,z) components and $\epsilon$  is the totally antisymmetric tensor. The bond operators satisfy bosonic commutation relations  $\left[s_{i},s^{\dagger}_{i}\right]=1$, $\left[t_{i,\alpha},t^{\dagger}_{i,\beta}\right]=\delta_{\alpha,\beta}$ and  $\left[s_{i},t^{\dagger}_{i,\alpha}\right]=1$. 
We will calculate the one particle boson Green's function
using Feynman diagrams for triplet operators and find the excitation spectrum. 
Our calculations are for zero temperature. 
In order to ensure that the 
physical states are either singlets or triplets one has to impose 
the constraint $s^{\dag} s+ \sum_{\alpha}t_{\alpha}^{\dag} t_{\alpha} = 1$  on every bond where 
$s$ (singlet) and $t_{\alpha}$ (triplet) are bond operators. 

In Refs.~\onlinecite{langari,thalmeier} this has been implemented on a mean field level by introducing a chemical potential as Lagrange parameter. In this approach average amplitudes $\bar{s}=\langle s_i\rangle$ and $\bar{t}=\langle{t_{i\alpha}\rangle}$ are introduced  and their self consistent solutions are found by minimizing the total ground state energy. Here $\bar{t}\neq 0$ denotes a triplet condensed state with magnetic order. The chemical potential adjusts itself such that the averge constraint $\bar{s}^2=1-\bar{t}^2$ is approximately satisfied. In fact in the paramagnetic region $(\bar{t}=0)$ it was found \cite{langari} that $\bar{s}$ is only a few per cent below singlet saturation $\bar{s}=1$ even close to the QCP where triplet excitations become soft. The zero point energy of the latter contribute to the ground state energy. Since in the mean field approach the number of triplet bosons on a given bond is not constrained there are contributions from unphysical states in the ground state energy.

In the present work we are therefore using a more advanced implementation of the local constraint which can be  written as $s_i^\dagger s_i=(1-\sum_\alpha t^\dagger_{i\alpha}t_{i\alpha})$. It may be satisfied if either $s^\dagger_i s_i=1$ and $\sum_\alpha t^\dagger_{i\alpha}t_{i\alpha}=0 $  or  $s^\dagger_i s_i=0$ and $\sum_\alpha t^\dagger_{i\alpha}t_{i\alpha}=1 $. To project out unphysical states on every bond with more than one excited triplet  one has to require $\sum_{\alpha}t^\dagger_{i\alpha}t^\dagger_{i\alpha}=0$. This may be achieved by introducing an on-site repulsion U of triplet bosons \cite{kotov} which is then taken in the hard core limit $U\rightarrow\infty$, see  Eq.~(\ref{e50}) below. As starting point for noninteracting triplets we use the unconstrained case with $s\rightarrow \bar{s}=1$ in Eq.~(\ref{e5}). In the paramagnetic case which we consider here this is well justified by the mean field result mentioned above. In principle one might think of a combined approach keeping $\bar{s}$ as a variational parameter within the hard core boson approximation. We will discuss this further in Sect.~\ref{sect:disc}.
This hard core boson approach can be applied to any
model, for which the excitations in the disordered phase are
triplets above a strong coupling singlet ground state. 
The Hamiltonian in Eq.~(\ref{e10}) has three control parameters, $J, \delta$ and $\Delta$. Using
the bond operator transformations  in the Kondo-necklace model of Eq.~(\ref{e10}), we obtain the effective Hamiltonian 
\begin{equation}
H=H_{2}+H_{3}+H_{4},
 \label{e20}
\end{equation}
where $H_{2}$ is the one particle part of the Hamiltonian. It is composed of two terms
\begin{equation}
H_{2}=H_{J_{\perp}}+H_{1},
\label{e25}  
\end{equation}
where the exchange term $H_{J_{\perp}}$ is diagonal in terms of the bond operators 
and $H_{1}$ has pairing terms between boson triplets which results in non-conservation of the triplet bosons and the possible formation of a Bose-Einstein condensate  of triplet bosons describing the magnetically ordered state. This term 
leads to a nonzero anomalous expectation value $ \left\langle  t_{\alpha} t_{\alpha} \right\rangle $ or corresponding anomalous Green's function. In terms of bond operators,
$H_{J_{\perp}}$ and $H_{1} $ 
are given by 
\begin{equation}
H_{J_{\perp}}=J_{\perp}\sum_{i}\bigl(\frac{(1+\Delta)}{2}
[t^{\dag}_{i,x}t_{i,x}+t^{\dag}_{i,y}t_{i,y}]+t^{\dag}_{i,z}t_{i,z}\bigr)\;,
\label{e30}
\end{equation}
\begin{equation}
H_{1}=\frac{J}{4} 
 \sum_{\left\langle i,j\right\rangle }\sum_{\alpha=x,y}(t_{i,\alpha}(t_{j,\alpha}+t_{j,\alpha}^{\dagger})+h.c.)\nonumber\\
+\frac{J\delta}{4}\sum_{\left\langle i,j\right\rangle}(t_{i,z}(t_{j,z}+t_{j,z}^{\dagger})+h.c.)\;.
 \label{e35}
\end{equation} 
The other parts of the Hamiltonian which describe triplet boson interactions are represented by
\begin{eqnarray}
H_{3}&=&\frac{J}{4}\sum_{\left\langle i,j\right\rangle }\Big( i[(t_{i,x}+t^{\dagger}_{i,x})(t^{\dagger}_{j,y}t_{j,z}-t^{\dagger}_{j,z}t_{j,y})+(t_{i,y}+t^{\dagger}_{i,y})(t^{\dagger}_{j,z}t_{j,x}-t^{\dagger}_{j,x}t_{j,z}) 
\nonumber\\
&+&\delta(t_{i,z}+t^{\dagger}_{i,z})(t^{\dagger}_{j,x}t_{j,y}-t^{\dagger}_{j,y}t_{j,x})]+h.c.\Big),
\end{eqnarray}
\begin{eqnarray}
H_{4}&=&-\frac{J}{4}\sum_{\langle i,j \rangle
}\Big((t^{\dag}_{i,y}t_{i,z}-h.c.)(t^{\dagger}_{j,y}t_{j,z}-h.c.)+(t^{\dagger}_{i,x}t_{i,z}-h.c.)(t^{\dagger}_{j,x}t_{j,y}-h.c.) \nonumber\\
&+&\delta(t^{\dagger}_{i,x}t_{i,y}-h.c.)(t^{\dagger}_{j,x}t_{j,y}-h.c.)\Big).
\label{e40}
\end{eqnarray}
The Hamiltonian can be written in terms of triplet Fourier components, $t_{i,\alpha}=\frac{1}{\sqrt{N}} 
\sum_{\textbf{k},\alpha}t_{k,\alpha}e^{i\overrightarrow{k}.\overrightarrow{R_{i}}}$, leading to the quadratic form

\begin{equation}
H_{2}=\sum_{k,\alpha=x,y,z}A_{k,\alpha}t^{\dagger}_{k,\alpha}t_{k,\alpha}
+\sum_{k,\alpha=x,y,z}\frac{B_{k,\alpha}}{2}(t^{\dagger}_{k,\alpha}t^{\dagger}_{-k,\alpha}+h.c.).
\end{equation}
The coefficients in the above equation are
\begin{eqnarray}
A_{k,z}=J_{\perp}+\delta J\xi_{k}\hspace{3mm}&,&\hspace{3mm}
A_{k,(x,y)}=\frac{J_{\perp}}{2}(1+\Delta)+ J\xi_{k},\nonumber\\
B_{k,z}=\delta J\xi_{k}\hspace{3mm}&,&\hspace{3mm}B_{k,(x,y)}= J\xi_{k},\nonumber\\
\xi_{k}=[\cos k_{x}&+&\cos k_{y}]/2.
\end{eqnarray}
Also for $H_{3}$ we obtain
\begin{eqnarray}
H_{3}&=&iJ\sum_{k_{1},k_{2},k_{3}=k_{1}+k_{2}}\xi_{k_{1}}(t^{\dagger}_{x,k_{1}}t^{\dagger}_{y,k_{2}}t_{z,k_{3}}-t^{\dagger}_{x,k_{1}}t^{\dagger}_{z,k_{2}}t_{y,k_{3}}+t^{\dagger}_{y,k_{1}}t^{\dagger}_{z,k_{2}}t_{x,k_{3}}\nonumber\\
&-&t^{\dagger}_{y,k_{1}}t^{\dagger}_{x,k_{2}}t_{z,k_{3}}+\delta t^{\dagger}_{z,k_{1}}t^{\dagger}_{x,k_{2}}t_{y,k_{3}}-\delta t^{\dagger}_{z,k_{1}}t^{\dagger}_{y,k_{2}}t_{x,k_{3}}) .
\end{eqnarray}
Because $H_{3}$ and $H_{4}$  are of higher order in triplet operators
they will lead to only small corrections in the spectrum. Therefore the effect
 of $H_{3}$ and $H_{4}$ may be taken into account on a mean field level.

The dominant
contribution to the renormalization of the spectrum comes from
the constraint where only one of the triplet states can be excited
on every site (the hard-core condition) 
$t^{\dag}_{\alpha i}t^{\dag}_{\beta i}=0$,
which can be taken into account by introducing an
infinite on-site repulsion between the bosons
\begin{equation}
H_{U}=U\sum_{i,\alpha,\beta}t^{\dag}_{\alpha i}t^{\dag}_{\beta i
}t_{\beta i}t_{\alpha i} ,  \hspace{5mm} U\longrightarrow\infty.
\label{e50}
\end{equation}
Writing $H_{U}$ in terms of Fourier transforms of boson operators we obtain  
\begin{equation}
H_{U}=U\sum_{k,k',q,\alpha,\beta}t^{\dag}_{\alpha
k+q}t^{\dag}_{\beta k'-q }t_{\beta k'}t_{\alpha k} \;.
\label{e60}
\end{equation}


\section{Green's function formalism in the Bosonic triplet gas}
The second part of  $H_{2}$ leads to the
non-interacting normal Green's function and in addition to anomalous
Green's function. Therefore, we introduce the single particle
Green's function for the non-interacting Hamiltonian which
help us to obtain the 
interacting ($H_{U}$) Green's function for the triplet operators  by using
Dyson's equation.
Implementing the  Bogoliubov
transformation $t_{k,\alpha}=u_{k,\alpha}\tilde{t}_{k,\alpha}+v_{k,\alpha}
\tilde{t}^{\dag}_{-k,\alpha}$ we obtain $\omega^{2}_{k,\alpha}=A^{2}_{k,\alpha}-B^{2}_{k,\alpha}$ for the excitation spectrum at
the quadratic level ($H_{2}$ only).
The Bogoliubov coefficients
are $u^{{2}}_{k,\alpha} (v^{2}_{k,\alpha})=(-)\frac{1}{2}+\frac{A_{k,\alpha}}{2\omega_{k,\alpha}}$.
The non-interacting normal triplet Green's
function is $G^{n}_{\alpha}(k,t)=-i\langle T(t_{k,\alpha}(t)
t^{\dag}_{k,\alpha}(0))\rangle$ and the anomalous Green's function
is $G^{a}_{\alpha}(k,t)=-i\langle T(t^{\dag}_{k,\alpha}(t)
t^{\dag}_{-k,\alpha}(0))\rangle$. Together we have
\begin{equation}
G^{n}_{\alpha}(k,\omega)=\frac{u_{k,\alpha}^{2}}{\omega-\omega_{k,\alpha}+i\eta}-
\frac{v_{k,\alpha}^{2}}{\omega+\omega_{k,\alpha}-i\eta},
\label{e100}
\end{equation}
\begin{equation}
G^{a}_{\alpha}(k,\omega)=\frac{u_{k,\alpha}
v_{k,\alpha}}{\omega-\omega_{k,\alpha}+i\eta}-
\frac{v_{k,\alpha}u_{k,\alpha}}{\omega+\omega_{k,\alpha}-i\eta}.
\label{e110}
\end{equation}
The interacting Green's functions are obtained from Dyson's equation for each Green's
function (anomalous or normal).
The perturbation expansion for the interacting Green's functions (for each polarization component of 
the triplet bosons) is written by
\begin{equation}
\overline{G}(k,\omega)=\overline{G^{0}}(k,\omega)
(1-\overline{G^{0}}(k,\omega)\overline{\Sigma}(k,\omega))^{-1}.
\label{e250}
\end{equation}
The interacting Green's function ($\overline{G}(k,\omega)$) and the
self-energy ($\overline{\Sigma}(k,\omega)$) are $2\times2$ matrices
\begin{eqnarray}
\overline{G}(k,\omega)= \left(
                              \begin{array}{cc}
                                G_{n}(k,\omega) & G_{a}(k,\omega) \\
                                G_{a}(k,\omega) & G_{n}(-k,-\omega) \\ 

  \end{array}
\right),\;\;\;
\overline{\Sigma}(k,\omega)= \left(
                              \begin{array}{cc}
                                \Sigma_{n}(k,\omega) & \Sigma_{a}(k,\omega) \\
                                \Sigma_{a}(k,\omega) & \Sigma_{n}(-k,-\omega) \\

  \end{array}
\right).
\label{e255}
\end{eqnarray}
The matrix form of Green's function can be simply expressed by 
 $\overline{G}(k,t)=\langle T( \Phi(k,t) \Phi^{\dag}(k,0))\rangle$, where
 $\Phi^{\dag}(k,t)=(t^{\dag}(k,t) \;\;\; t(-k,t))$ is the row vector.
Inserting the elements of Eq.(\ref{e100}), Eq.(\ref{e110}) into Eq.(\ref{e250}), the normal and anomalous
interacting Green's function will be obtained by 
\begin{eqnarray}
&&G_{n,\alpha}(k,\omega)=\frac{\omega+A_{k,\alpha}+\Sigma_{n,\alpha}(-k,-\omega)}{[\omega+A_{k,\alpha}+\Sigma_{n,\alpha}(k,-\omega)][\omega-A_{k,\alpha}-\Sigma_{n,\alpha}(k,\omega)]+(B_{k}+\Sigma_{a,\alpha}(k,\omega))^{2}},\nonumber\\
&&G_{a,\alpha}(k,\omega)=\frac{B_{k,\alpha}+\Sigma_{a,\alpha}(k,\omega)}{[\omega+A_{k,\alpha}+\Sigma_{n,\alpha}(k,-\omega)][\omega-A_{k,\alpha}-\Sigma_{n,\alpha}(k,\omega)]+(B_{k}+\Sigma_{a,\alpha}(k,\omega))^{2}}.
\label{e270}
\end{eqnarray}
The normal and anomalous self-energy are due to boson interactions $H_{3}$ and $H_{U}$ and will be discussed in the next sections. 
Here we are interested to find the one particle excitations which are the poles of
the normal triplet Green's function. The Green's function should be separated
into the bosonic excitation contribution and incoherent background (including collective modes).  
Indeed, poles of the one particle Green's function of the triplet bosons result in
low energy excitations of the Hamiltonian which vanish close to the 
critical point. To get the single particle excitation the self-energy
is expanded for low energies leading to
\begin{eqnarray}
&&G_{n,\alpha}(k,\omega)=\frac{\omega+A_{k,\alpha}+\Sigma_{n,\alpha}(k,0)-\omega\partial_{\omega}\Sigma_{n,\alpha}(k,0)}
{D+(B_{k,\alpha}+\Sigma_{a,\alpha}(k,0))^{2}},\nonumber \\
&&D\equiv D_1 \cdot D_2, \nonumber \\
&&D_1\equiv[\omega+A_{k,\alpha}+\Sigma_{n,\alpha}(k,0)-\omega\partial_{\omega}\Sigma_{n,\alpha}(k,0)],\nonumber \\
&&D_2\equiv[\omega-A_{k,\alpha}-\Sigma_{n,\alpha}(k,0)-\omega\partial_{\omega}\Sigma_{n,\alpha}(k,0)].
\label{e320}
\end{eqnarray}
Splitting  Eq.(\ref{e320}) into partial fractions leads to the single particle ($sp$) parts
\begin{equation}
G_{n,\alpha}^{sp}(k,\omega)=
\frac{Z_{k,\alpha}U_{k,\alpha}^{2}}{\omega-\Omega_{k,\alpha}+i\eta}-\frac{Z_{k,\alpha}V_{k,\alpha}^{2}}{\omega+\Omega_{k,\alpha}-i\eta},
\label{e330}
\end{equation}
where the renormalized triplet spectrum and the renormalized single particle weight constants are given by
\begin{eqnarray}
\Omega_{k,\alpha}&=&Z_{k,\alpha}\sqrt{[A_{k,\alpha}+\Sigma_{n,\alpha}(k,0)]^{2}-[B_{k,\alpha}+\Sigma_{a,\alpha}(k,0)]^{2}}\nonumber\\
&&Z_{k,\alpha}^{-1}=1-(\frac{\partial \Sigma_{n,\alpha}}{\partial \omega})_{\omega=0}\nonumber\\
&&U_{k,\alpha}^{2} (V_{k,\alpha}^{2})=(-)\frac{1}{2}+\frac{Z_{k,\alpha}[A_{k,\alpha}+\Sigma_{n,\alpha}(k,0)]}{2\Omega_{k,\alpha}}.
\label{e340}
\end{eqnarray}
The renormalized weight constant is indeed the residue of the single particle pole in the
Green's function. For the non-interacting system it is equal to one. 

\section{Calculation of boson self-energy due to $H_{U}$ and $H_{3}$ }

Since the Hamiltonian $H_{U}$ in Eq.(\ref{e60}) is short ranged and U  is large, the ladder diagram
approach \cite{fetter} may be applied. 
This approach is suitable to solve Dyson's equation in order
to get the boson Green's function. Formally this is quite similar to Ref.~\onlinecite{kotov}, however, technically
more demanding due to the effect of anisotropies $(\delta,\Delta)$.\\

\begin{figure}
\includegraphics[width=12cm]{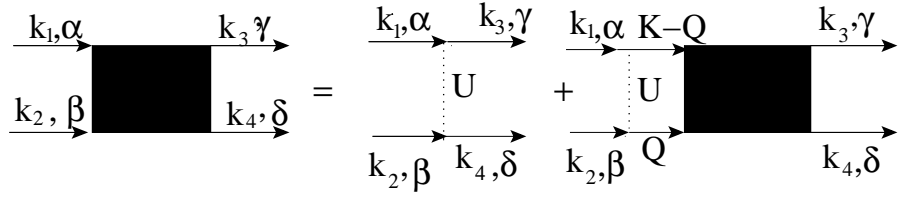}
\caption{\label{fig1} Ladder diagram for the triplet boson scattering amplitude 
$\Gamma_{\alpha\beta,\gamma\delta}(k_{1},k_{2};k_{3},k_{4})$.}
\end{figure}

\begin{figure}[ht]
\includegraphics[width=12cm]{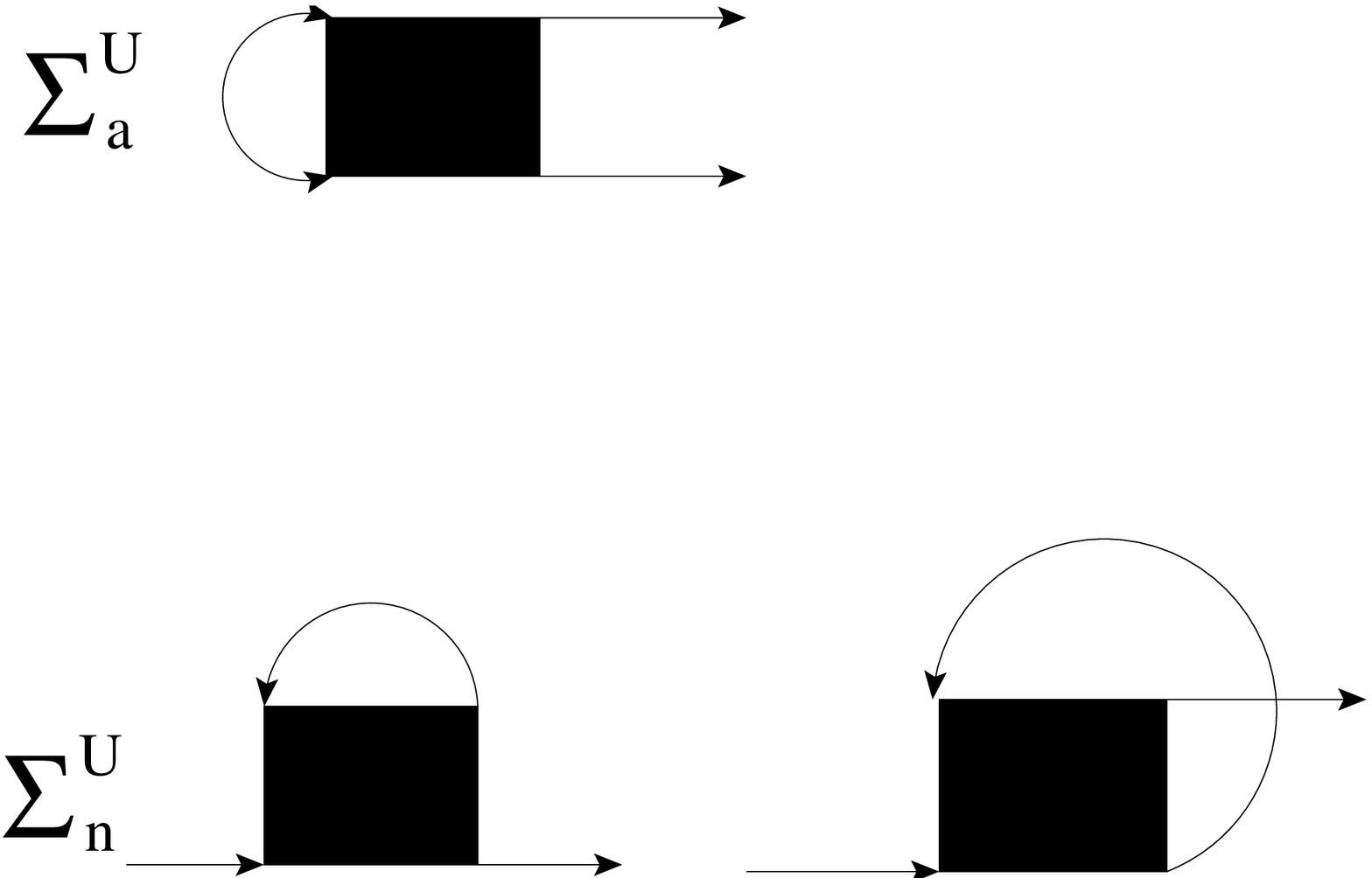}
\caption{\label{fig2} Diagrams for the normal and anomalous single particle self-energy due to hard core term $H_U$ and using  the two-particle scattering amplitude in Fig.~\ref{fig1}.}
\end{figure}

Now, we should impose the hard core repulsion due to the
Hamiltonian $H_U$ and obtain the interacting normal Green's function by
Dyson's equation. Firstly, we introduce the scattering amplitude
$\Gamma_{\alpha\beta,\gamma\delta}(k_{1},k_{2};k_{3},k_{4})$ of triplet bosons where
$k_{i}=(\overrightarrow{k_{i}},k^{0}_{i})$.  The ladder approximation
satisfies a Bethe-Salpeter equation which is shown in Fig.~\ref{fig1} and
written in Eq.(\ref{e120}).
The scattering amplitude or self-energy for the two particle Green's function
depends on the total energy and
momentum  of the incoming particles
$\overrightarrow{K}=\overrightarrow{p_{1}}+\overrightarrow{p_{2}}$.
The non-retarded and local character of $U$ leads to
$\Gamma_{\alpha\beta,\gamma\delta}=\Gamma\delta_{\alpha\gamma}\delta_{\beta\delta}$.
The basic approximation made in the derivation of $\Gamma(K)$ is
that we neglect all anomalous scattering vertices, which are
present in the theory due to existence of anomalous Green's functions.
For the scattering amplitude shown in Fig.~\ref{fig1}, according to the
Feynman rules in momentum space we can write (note
$p\equiv(p_{0},\overrightarrow{p})$ )
\begin{eqnarray}
&&\Gamma_{\alpha\beta,\alpha\beta}(p_{1}p_{2};p_{3}p_{4})=U(p_{1}-p_{3}) \nonumber \\
&&+i
(2\pi)^{-4}\int \Big[ d^{4}Q U(Q-p_{2})
G^{0}_{\alpha\alpha}(Q)G^{0}_{\beta\beta}(p_{1}+p_{2}-Q) 
\Gamma_{\alpha\beta,\alpha\beta}(p_{1}+p_{2}-Q,Q;p_{3}p_{4}) \Big].
 \label{e120}
\end{eqnarray}
In the above equation $U$ is independent of momentum and energy.
Consequently the $\Gamma$-function depends only on the  sum of the incoming
momentum and energy, and  does not depend separately on the  momentum and energy
of the incoming particles. Therefore, $p_{1}+p_{2}=p_{3}+p_{4}\equiv
K=(\overrightarrow{K},\omega)$ which simplifies Eq.(\ref{e120}) to
\begin{equation}
\Gamma_{\alpha\beta,\alpha\beta}(\overrightarrow{K},\omega)=U+i(2\pi)^{-4}\int
d^{4}Q U G_{\alpha\alpha}^{0}(Q)G_{\beta\beta}^{0}(K-Q)\Gamma_{\alpha\beta,\alpha\beta}(K,\omega).
\label{e130}
\end{equation}
However, the key observation is that all anomalous contributions
are suppressed by an additional small parameter present in the
theory - the density of the triplet excitation
$n_{i}=\sum_{\alpha}\langle t^{\dag}_{\alpha i}t_{\alpha
i}\rangle=N^{-1}\sum_{\overrightarrow{q},\alpha}v_{q,\alpha}^{2}\approx 0.1 $.
Indeed, both terms of the anomalous scattering matrix are proportional to 
$v_{q,\alpha}^{2}$ which are neglected. 
By replacing the  noninteracting normal Green's function (Eq.(\ref{e100}))
in the Bethe-Salpeter equation (Eq.(\ref{e130})) and taking the limit $U\longrightarrow\infty$ we obtain 
the scattering matrix in the form  (see Appendix A)
\begin{equation}
\Gamma_{\alpha\beta,\alpha\beta}(\overrightarrow{K},\omega)=-\Big(\frac{1}{N}
\sum_{\overrightarrow{q}}
\frac{u_{q,\alpha}^{2}u_{K-q,\beta}^{2}}{\omega-\omega_{q,\alpha}-\omega_{K-q,\beta}}-
\frac{v_{q,\alpha}^{2}v_{K-q,\beta}^{2}}{\omega+\omega_{q,\alpha}+\omega_{K-q,\beta}}
\Big)^{-1}.
\label{e170}
\end{equation}
Now, we can calculate the single particle self-energy (Fig \ref{fig2}) of 
bosons by utilizing the two particle self-energy ($\Gamma$) shown in Fig.~\ref{fig1} and 
obtained in Eq.(\ref{e170})\cite{fetter}. Because of the
strong interaction between the triplet bosons we should carry out the expansion in
Dyson's equation to infinite order. Therefore, $\Sigma^{U}_{\alpha\alpha}(k)$ is written by 
\begin{equation}
\Sigma^{U}_{\alpha\alpha}(k)=\sum_{\gamma\beta}\int^{\infty}_{-\infty}d^{4}p\Gamma_{\alpha\beta,\gamma\delta}(p,k;k,p)G^{0}_{\gamma\beta}(p)+
\sum_{\beta\delta}\int^{\infty}_{-\infty}d^{4}p\Gamma_{\alpha\beta,\gamma\delta}(p,k;p,k)G^{0}_{\delta\beta}(p). \hspace{3mm}
\label{e190}
\end{equation}
(Note that
$\Gamma_{\alpha\beta,\gamma\delta}=\Gamma_{\alpha\beta,\alpha\beta}\delta_{\alpha\gamma}\delta_{\beta\delta}$
and
$U$ is frequency and momentum independent). For example, the x-component of the self-energy is written by
\begin{eqnarray}
\Sigma^{U}_{xx}(k)&=&2(\frac{i}{2\pi})^{4}\int^{\infty}_{-\infty}d^{4}p\Gamma_{xx,xx}(p+k)G_{xx}^{0}(p)
+(\frac{i}{2\pi})^{4}\int^{\infty}_{-\infty}d^{4}p\Gamma_{xz,xz}(p+k)G_{zz}^{0}(p)\nonumber\\
&&+(\frac{i}{2\pi})^{4}\int^{\infty}_{-\infty}d^{4}p\Gamma_{xy,xy}(p+k)G_{yy}^{0}(p).
\label{e200}
\end{eqnarray}
In the above equation the first and third terms are similar. 
We integrate over the internal energy in complex plane on a
contour in the upper half plane since $G^{0}(p)$ (Eq.(\ref{e200})) is anti-time ordered. Consequently, $\Sigma^{U}_{xx}(k,k_{0}\equiv\omega)$ will be written by (N is the 
number of cells in the  lattice)
\begin{eqnarray}
\Sigma^{U}_{xx}(k,k_{0}\equiv\omega)&=&3\frac{i}{2\pi N}\sum_{p}\int^{\infty}_{-\infty}dp_{0}\Gamma_{xx,xx}(p+k,p_{0}+k_{0})\frac{-v_{p,x}^{2}}{p_{0}+\omega_{p,x}-i\eta}\nonumber\\
&+&\frac{i}{2\pi N}\sum_{p}\int^{\infty}_{-\infty}dp_{0}\Gamma_{xz,xz}(p+k,p_{0}+k_{0})\frac{-v_{p,z}^{2}}{p_{0}+\omega_{p,z}-i\eta}\nonumber\\
&=&\frac{3}{N}\sum_{p}v_{p,x}^{2}\Gamma_{xx,xx}(p+k,\omega-\omega_{p,x})
+\frac{1}{N}\sum_{p}v_{p,z}^{2}\Gamma_{xz,xz}(p+k,\omega-\omega_{p,z}).
\label{e210}
\end{eqnarray}
In the dilute gas approximation there are other diagrams which are formally at most linear in $n_{t}$ (density of triplet bosons) but still numerically give contributions much smaller than Eq.~(\ref{e210}). We should also consider the anomalous self-energy related to $H_{U}$ which is obtained from the vertex function in Eq.(\ref{e170}). The anomalous self-energy (non-diagonal elements in the Eq.(\ref{e255})) is written as (Fig \ref{fig2})
\begin {equation}
 \Sigma^{U}_{A}=\frac{1}{N}\sum_{q}u_{q}v_{q}\Gamma(0,0).
\end {equation}
The self-energy is proportional to $\sum_{q}u_{q}v_{q}$ since the vertex function
is independent of $k$. It is then proportional to the anomalous Green's function.
It is negligible as compared with the normal self-energy (diagonal parts).\\

\begin{figure}[ht]
\includegraphics[width=12cm]{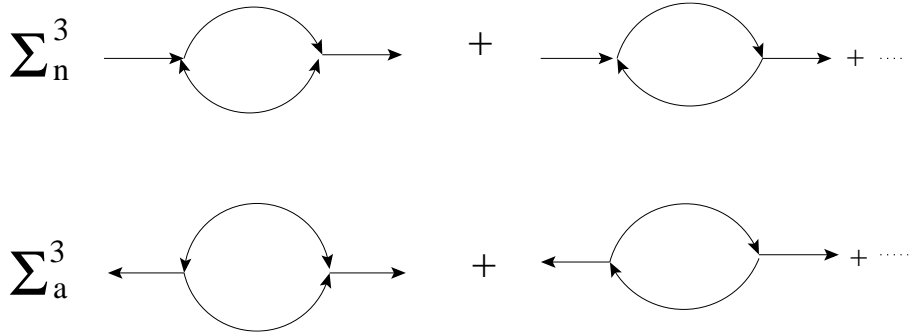}
\caption{\label{fig3} One loop diagrams for normal and anomalous self-energy arising from the three point interaction H$_3$.}
\end{figure}

We now consider the $H_{3}$ contribution in the normal and anomalous parts 
of the self-energy. The normal part should be added to the self-energy due to $H_{U}$. 
Since $H_{3}$ is much weaker than $H_{U}$ it is sufficient 
to obtain the second order perturbation result of Dyson's series for each
component of the normal Green's function.
The formula for the self-energy contribution (either anomalous or normal) is quite lengthy and has been presented in
Appendix B and the corresponding Feymann diagrams are shown in Fig.~\ref{fig3}.
The normal self-energy contribution of  $H_{3}$ is proportional to $u^{4}$ which
therefore dominates 	the anomalous one.


\section{Effect of $H_{4}$ on the renormalization of the spectrum}
Because $H_4$ is composed of quartic terms in the triplet operators its effect
should be very small. It is therefore treated in mean field approximation by contracting the quartic operator products into all possible pairs. This is equivalent to take only 
the one loop diagrams (first order in $J$) into account. On mean field level we have 
$O_1 O_2=\left\langle O_1 \right\rangle O_2+\left\langle O_2 \right\rangle O_1-
\left\langle O_2 \right\rangle \left\langle O_1\right\rangle$ where each $O_1$ and $O_2$ is a pair of 
boson triplet operator. We can write for each pair of operators
\begin{eqnarray}
 &&\left\langle t_{i,\alpha}^{\dagger}t_{j,\alpha}\right\rangle =\frac{i}{2\pi N}\int _{-\infty}^{\infty} d\omega \sum_{k}e^{ik.(R_{j}-R_{i})-i\omega 0^{+}}G_{n}^{\alpha\alpha}(k,\omega)
=\frac{1}{N}\sum_{k}e^{ik.(R_{j}-R_{i})}v_{k,\alpha}^{2},\nonumber\\
&&\left\langle t_{i,\alpha}^{\dagger}t_{j,\alpha}^{\dagger}\right\rangle=\frac{1}{N}\sum_{k}e^{ik.(R_{j}-R_{i})}u_{k,\alpha}v_{k,\alpha}.
\label{e410}
\end{eqnarray}
Thus, the effect of $H_{4}$ is to renormalize  $A$ and $B$ coefficients defined in 
$H_2$ in the form
\begin{eqnarray}
 &A_{k,z}&\longrightarrow A_{k,z}+2J\xi_{k}\frac{1}{N}\sum_{q}(v_{q,x}^{2})\xi_{q},\nonumber\\
&B_{k,z}&\longrightarrow B_{k,z}-2J\xi_{k}\frac{1}{N}\sum_{q}(u_{q,x}v_{q,x})\xi_{q},\nonumber\\
&A_{k,(x,y)}&\longrightarrow A_{k,(x,y)}+J\xi_{k}\frac{1}{N}\sum_{q}(\delta v_{q,(x,y)}^{2}+v_{q,z}^{2})\xi_{q},\nonumber\\
&B_{k,(x,y)}&\longrightarrow B_{k,(x,y)}-J\xi_{k}\frac{1}{N}\sum_{q}(\delta u_{q,(x,y)}v_{q,(x,y)}+u_{q,z}v_{q,z})\xi_{q}.
\label{e500}
\end{eqnarray}       
The renormalized coefficients (Eq.(\ref{e500})) will be considered to calculate
the normal and anomalous self-energy which are independent of energy 
(nonretarded in time representation). The self-consistent solution of Eqs.(\ref{e170},\ref{e210},\ref{e280},\ref{e290},\ref{e300},\ref{e310},\ref{e500},\ref{e340}) describes the quantum critical behaviour of this model which will be discussed in the
following sections.



\section{The quantum critical point and the gap exponent}
Close to the critical point quantum fluctuations exist over all length scales which
define a scaling behavior for the physical quantities. The correlation length scales like
$\xi\sim |J_\perp - J_{\perp c}|^\nu$ where $\nu$ is a critical exponent. 
This is related to the scaling behavior of the excitation gap in the Kondo singlet phase which vanishes like
\begin{equation}
 E_g \sim |J_{\bot}-J_{\bot c}|^{\phi},
\label{e551}
\end{equation}
as $J_{\bot}$ approaches its critical value $J_{\bot c}$. Here
$\phi$ is called the gap exponent which is connected to the universality class of the
quantum critical point. From general scaling arguments one expects $\phi=\nu z$ where $z$ is the dynamical critical exponent that determines the effective dimension $D_{eff}$ of the model at 
T=0 according to $D_{eff}=D+z$.
The spin excitation  gap is defined by the energy of
triplet excitations with x-polarization close to the antiferromagnetic wave vector ${q}_{0}$.
In its vicinity  ($|k-q_0|\ll 1$)
the triplet dispersion can be approximated by
\begin{eqnarray}
\omega_{k,x}=\sqrt{E^{2}_{g}+c_{x}^{2}(k-q_{0})^{2}},
\label{e600}
\end{eqnarray}
where $c_{x}$ is the spin-wave velocity \cite{weihong,kotov,hida}  and
$q_{0}=(\pi,\pi)$. 
There is no analytical expression for the spectrum of excitations, 
therefore we use numerical results to get the spin wave velocity. 
The slope of dispersion of the x-component excitations close to $q_{0}$ 
is calculated numerically which is the spin wave velocity $c_x$.
To find the energy gap we should consider the excitation energy at the wave vector $q_{0}$
\begin{eqnarray}
 E^{2}_{g}=Z_{q_{0},x}(A_{q_{0},x}^{2}-B_{q_{0},x}^{2}),
\label{e650}
\end{eqnarray}
where the renormalized constants have been obtained in the previous section
\begin{eqnarray}
 &&A_{k,x}=\frac{J_{\perp}}{2}(1+\Delta)+ J\xi_{k}+\Sigma_{n,x}^{U}(k)+
\Sigma_{n,x}^{3}(k)+\frac{J \delta \xi_{k}}{N}\sum_{q}( Z_{q,x}v_{q,x}^{2}+Z_{q,z}v_{q,z}^{2})\xi_{q}, \nonumber\\
&&B_{k,x}= J\xi_{k}+\Sigma_{a,x}^{U}(k)+\Sigma_{a,x}^{3}(k)-
\frac{J \xi_{k}}{N}\sum_{q}(\delta u_{q,x}v_{q,x}+u_{q,z}v_{q,z})\xi_{q}.
\label{e550}
\end{eqnarray}
For the values of $A_{q_{0},x}$ and $B_{q_{0},x}$ at the critical point,
$A_{q_{0},x}^{c}=-B_{q_{0},x}^{c}$ holds.

The energy gap in Eq.~(\ref{e650}) vanishes at the quantum critical point ($J_{\bot c}$)
and its behavior close to this point defines the scaling in Eq.(\ref{e551}).
We now look for the variation of the energy gap as $J_{\bot}$ deviates from $J_{\bot c}$
which is given by the variation of $A_{q_{0},x}$ and $B_{q_{0},x}$ with respect 
to $J_{\bot}$ deviation.
The deviation of $J_{\bot}$ from the critical point is defined by 
$\delta_{s}J_{\bot}\equiv J_{\bot}-J_{\bot c}$. 
Therefore, close to critical point  $A_{q_{0},x}$ and $B_{q_{0},x}$ 
can be written as
 \begin{eqnarray}
&&A_{q_{0},x}=A_{q_{0},x}^{c}+\frac{1+E_g}{2}\delta_{s} J_{\perp}+\partial \Sigma_{n,x}^{U}(\pi,\pi)+\partial \Sigma_{n,x}^{3}(\pi,\pi)-\frac{J \delta}{N}\sum_{q}( Z_{q,x}\partial v_{q,x}^{2})\xi_{q} ,\nonumber\\
&&B_{q_{0},x}=B_{q_{0},x}^{c}+\partial \Sigma_{a,x}^{3}(\pi,\pi)-\frac{J}{N}\sum_{q}( Z_{q,x}\partial u_{q,x}v_{q,x})\xi_{q},
  \label{e700}
 \end{eqnarray}
where $\partial X$ means the variation of $X$ with respect to $\delta_{s}J_{\bot}$.
If we substitute Eq.(\ref{e700}) into Eq.(\ref{e650}) and neglect terms quadratic in $E_g$ the variation of $A_{q_{0}}$ and $B_{q_{0}}$ must vanish
\begin{eqnarray}
 &&\partial A_{q_{0}}=\frac{1+E_g}{2}\delta_{s} J_{\perp}+\delta_{s}\Sigma_{n,x}^{U}(\pi,\pi)+\partial \Sigma_{n,x}^{3}(\pi,\pi)-
\frac{J \delta}{N}\sum_{q}(Z_{q,x}\partial v_{q,x}^{2})\xi_{q}=0 ,\nonumber\\
&&\partial B_{q_{0}}=\partial \Sigma_{a,x}^{3}(\pi,\pi)-\frac{J}{N}\sum_{q}(Z_{q,x}\partial u_{q,x}v_{q,x})\xi_{q}=0 .
\label{e850}
\end{eqnarray}
We now have to obtain the variation of each of the terms present in Eq.(\ref{e700}).
In the first step, we calculate the variation of the self-energy related to $H_{U}$ which is written by 
\begin{eqnarray}
\partial \Sigma^{U}_{x}(\pi,\pi)
&=&3\int \frac{d^{2}q}{(2\pi)^{2}}\partial v_{q,x}^{2}\Gamma_{xx,xx}(q+q_{0},-\omega_{q,x})+3\int \frac{d^{2}q}{(2\pi)^{2}}v_{q,x}^{2}\partial \Gamma_{xx,xx}(q+q_{0},-\omega_{q,x})\nonumber\\
&+&\int\frac{d^{2}q}{(2\pi)^{2}}\partial v_{q,z}^{2}\Gamma_{xz,xz}(q+q_{0},-\omega_{q,z})+\int\frac{d^{2}q}{(2\pi)^{2}} v_{q,z}^{2}\partial \Gamma_{xz,xz}(q+q_{0},-\omega_{q,z}),
\label{e750}
\end{eqnarray}
where $\partial v_{p,z}^{2}=0$. Indeed for $0\leq\delta < 1$ the z-component of the spectrum has a finite gap when 
the x-component becomes gapless at the quantum critical  point.
 The main contribution to the first integral in  Eq.~(\ref{e750}) comes from the 
small momenta ($q\sim E_g\ll1$) since
\begin{eqnarray}
\partial v_{q,x}^{2}=\frac{1}{2} \Big(\frac{\partial A_{q,x}}{\omega_{q,x}}+A_{q,x}\partial [\frac{1}{\omega_{q,x}}]\Big)\approx -\frac{A_{q_{0},x}^{c}E_g^{2}}{2(E_g^{2}+c_{x}^{2}(q-q_{0})^{2})^{3/2}},
\label{e800}
\end{eqnarray}
and according to Eq.(\ref{e850}) the variation $\partial A_{q,x}$ in this formula vanishes. 
We define the value of quantity $X$ at the critical point by $X^c$,
Taking into account the first correction to the triplet density $n_{b}$, the x-component of the vertex function can be written for small $q$ (see Ref.[\onlinecite{shevchenko}])
\begin{eqnarray}
 \Gamma_{xx,xx}(q,-\omega_{q,x})\approx\Gamma_{xx,xx}^{c}+\frac{{\Gamma_{xx,xx}^{c}}^{2}A_{0,x}^{c}}{4\pi c_{x}^{2}}\ln q,
\label{e900}
\end{eqnarray}
where $\Gamma_{xx,xx}(0,0)=\Gamma_{xx,xx}^{c}$.
The substitution of Eq.(\ref{e900}) in  Eq.(\ref{e750}) and replacing 
$q\longrightarrow E_g/J$ in the first integral of Eq.(\ref{e750})
leads to the following equation for the variation of $H_{U}$ self-energy
\begin{eqnarray}
 \partial \Sigma^{U}_{x}(\pi,\pi)=-\frac{3 A_{q_{0},x}^{c}E_g}{4\pi c_{x}^{2}}(\Gamma_{xx,xx}^{c}+\frac{{\Gamma_{xx,xx}^{c}}^{2}A_{0,x}^{c}}{4\pi c_{x}^{2}}\ln\frac{E_g}{J})+\frac{1}{3}\Gamma_{xz,xz}^{\prime}n_{b}\delta_{s} J_{\perp}+\Gamma_{xx,xx}^{\prime}n_{b}\delta_{s} J_{\perp}.
\label{e950}
\end{eqnarray}
In the second and fourth terms of Eq.(\ref{e750}) we have $\Gamma_{\alpha\beta,\alpha\beta}^{\prime}=\frac{\delta \Gamma_{\alpha\beta,\alpha\beta}(q,-\omega_{q})}{\delta J_{\bot}}$ and $n_{b}$ is the density of triplet excitations at the critical point. Now, we consider the effect of $H_{3}$ on the gap exponent. 
According to Appendix B the variation of self-energy contribution of $H_{3}$ is given by 
\begin{eqnarray}
\partial \Sigma_{n,x}^{3}(\pi,\pi)=
\frac{J^{2}}{2N}\Big(\chi u^c_{q=0,z}v^c_{q=0,z}+\varphi (v^c_{q=0,z})^{2}+\varphi (u^c_{q=0,z})^{2}\Big)\sum_{q}\partial (\frac{1+2v_{q,x}^{2}}{\omega_{q,z}+\omega_{q,x}}),
 \label{e1000}
\end{eqnarray}
where $\chi=-(4\delta+2+2\delta^{2}), \varphi=-(2\delta+1+\delta^{2})$. 
Since in the vicinity of critical point $\omega_{x,q_{0}}\ll\omega_{z,q_{0}}$ 
the variation of Eq.(\ref{e1000}) gives
\begin{eqnarray}
 \partial (\frac{1}{\omega_{q,x}+\omega_{q,z}})=-\frac{E_g^{2}A^{c}_{q_{0},x}}{4(E_g^{2}+c_{x}^{2}(q-q_{0})^{2})^{1/2}(\omega_{q,z}+\sqrt{E_g^{2}+c_{x}^{2}(q-q_{0})^{2}})^{2}}.
\label{e1010}
\end{eqnarray}
The integral of Eq.(\ref{e1010}) multiplied by $1+2v_{q,x}^{2}$ is 
proportional to  $E_g^{2}\ln E_g$ which can be neglected 
compared with the first term in Eq.(\ref{e950}).
Therefore, we can restrict to the variation of $1+2v_{q,x}^{2}$. 
Then, the dominant contribution of  Eq.(\ref{e1000}) is given by
\begin{eqnarray}
\partial \Sigma_{n,x}^{3}(\pi,\pi)=-\frac{A^{c}_{q_{0},x}J^{2}
\Big((u^c_{0,z}v^c_{0,z})\chi+((u^c_{0,z})^{2}+(v^c_{0,z})^{2})\varphi\Big)}{4\pi c_{x}^{2}\omega^c_{0,z}}E_g.
\label{e1020} 
\end{eqnarray}
We then calculate the last variation of the first expression in Eq.(\ref{e700}).
\begin{eqnarray}
 J\frac{1}{N}\sum_{q}(\delta Z_{q,x}\partial v_{q,x}^{2})\xi_{q}=-\frac{\delta Z_{0}JA^{c}_{q_{0},x}}{\pi c_{x}^{2}}E_g.
\label{e1030}
\end{eqnarray}
Let us define the following expressions
\begin{eqnarray}
&&\lambda \equiv \frac{3A^{c}_{q_{0},x}\Gamma_{xx,xx}^{c}}{4\pi c_{x}^{2}},\nonumber\\
&&\theta \equiv \frac{A^{c}_{q_{0},x}J^{2}\Big((u^c_{0,z}v^c_{0,z})\chi+((u^c_{0,z})^{2}+(v^c_{0,z})^{2})\varphi\Big)}{4\pi c_{x}^{2}\omega^c_{0,z}},\nonumber\\
&&\mu  \equiv \frac{\delta JA^{c}_{q_{0},x}}{\pi c_{x}^{2}},\nonumber\\
&&\sigma  \equiv \frac{1}{3}\Gamma_{xz,xz}^{\prime}n_{b}\delta J_{\perp}+\Gamma_{xx,xx}^{\prime}n_{b}\delta J_{\perp}.
\label{e1040}
\end{eqnarray}
From the substitution of Eqs.~(\ref{e1030},\ref{e1020},\ref{e950}) into the first expression of Eq.~(\ref{e850}) we get the following equation
\begin{eqnarray}
 E_g=\frac{(\frac{1+E_g}{2}+\sigma)\delta_{s}J_{\bot}}{\lambda+\theta+\mu}\Big(1-\frac{\lambda A^{c}_{q_{0},x}\Gamma_{xx,xx}^{c}}{4\pi c_{x}^{2}(\lambda+\theta+\mu)}ln\frac{\delta_{s}J_{\bot}}{J}\Big).
\label{e1050}
\end{eqnarray}
To find the gap exponent $\phi$ we should consider $E_g=(\delta_{s}J_{\bot})^{\phi}$. Indeed $\phi$ is the smallest exponent that can be considered for $E_g$. Finally  we obtain
\begin{eqnarray}
 \phi=1-\frac{\lambda A^{c}_{q_{0},x}\Gamma_{xx,xx}^{c}}{4\pi c_{x}^{2}(\lambda+\theta+\mu)}.
\label{e1060}
\end{eqnarray}
The last equation is obtained using $\omega_{x,q_{0}}\ll\omega_{z,q_{0}}$. 
For the isotropic case, $\omega_{x,q_{0}}=\omega_{z,q_{0}}\equiv \omega_{q_{0}}$, where the expression for $\Sigma^{U}_{x}(\pi,\pi)$ should be changed. The final result for the gap exponent of 
the isotropic case ($\delta=\Delta=1$) is given by
\begin{eqnarray}
\phi=1-\frac{\lambda A^{c}_{q_{0}}\Gamma^{c}}{4\pi c^{2}(\lambda+\mu+\theta)}.
 \label{e1070}
\end{eqnarray}
 In the above equation we have 
\begin{eqnarray}
 &&A^{c}_{q_{0},x}=A^{c}_{q_{0},z}\equiv A^{c}_{q_{0}},\nonumber\\
&&\lambda\equiv \frac{A^{c}_{q_{0}}\Gamma^{c}}{\pi c^{2}},\nonumber\\
&&\mu \equiv 2\frac{JA^{c}_{q_{0}}}{\pi c^{2}},\nonumber\\
&&\theta \equiv \frac{A^{c}_{q_{0}}J^{2}\Big(16(u^c_{0}v^c_{0})+8((u^c_{0})^{2}+(v^c_{0})^{2})\Big)}{4\pi c^{2}\omega^c_{0}}.
\label{e1080}
\end{eqnarray}
We have summarized the numerical values of the gap exponent for different anisotropies $(\delta,\Delta)$
in Tables \ref{t1} and \ref{t2}.


\section{The Quantum Critical Phase Diagrams, Numerical Results}

Our approach is based on the strong coupling limit, $J_{\bot}/J \rightarrow \infty$. In this limit the ground state has singlet character and a finite energy gap exists to the lowest excited triplet state. The increase of inter-site exchange
coupling ($J$) or decrease of on-site exchange ($J_\perp$) lowers the excitation gap which eventually vanishes. 
The position where the bosonic excitation gap vanishes defines the quantum critical 
point. At this point the condensation of a triplet takes place which induces the antiferromagnetic order.
The single particle excitation energy should be obtained in a self consistent 
solution of Eqs.(\ref{e170},\ref{e210},\ref{e280},\ref{e290},\ref{e300},\ref{e310},\ref{e500},\ref{e340}).
We should first replace $$u_{k,\alpha}\longrightarrow \sqrt{Z_{k,\alpha}}U_{k,\alpha}, v_{k,\alpha}\longrightarrow \sqrt{Z_{k,\alpha}}V_{k,\alpha}$$
in the self-energies of $H_{U},H_{3}$ and the renormalization expressions in Eq.(\ref{e500}). From an initial guess for $Z_{k,\alpha}, \Sigma_{n,\alpha}(k,0), \Sigma_{a,\alpha}(k,0)$
by using Eq.(\ref{e340}) we obtain corrected excitation energy and the renormalized Bogoliubov 
coefficients $(u, v)$. We repeat the procedure until the difference between the excitation energies 
in two consecutive steps is smaller than an acceptable error. The exponent of the gap is given by
Eq.(\ref{e1060}) which is calculated from the vertex function and spin wave velocity at the critical point and at the antiferromagnetic wave vector $Q=(\pi,\pi)$. 
We will discuss the numerical results of our calculations in the next subsections
for XY-case, i.e., $\delta=0$ and various sizes of on-site exchange anisotropy 
$0\leq\Delta\leq1$ and likewise  for $\Delta=1$  with various values of the inter-site exchange anisotropy $0\leq\delta\leq1$. 


\subsection{XY-case: $\delta=0$}
\begin{figure}[ht]
\includegraphics[width=8cm]{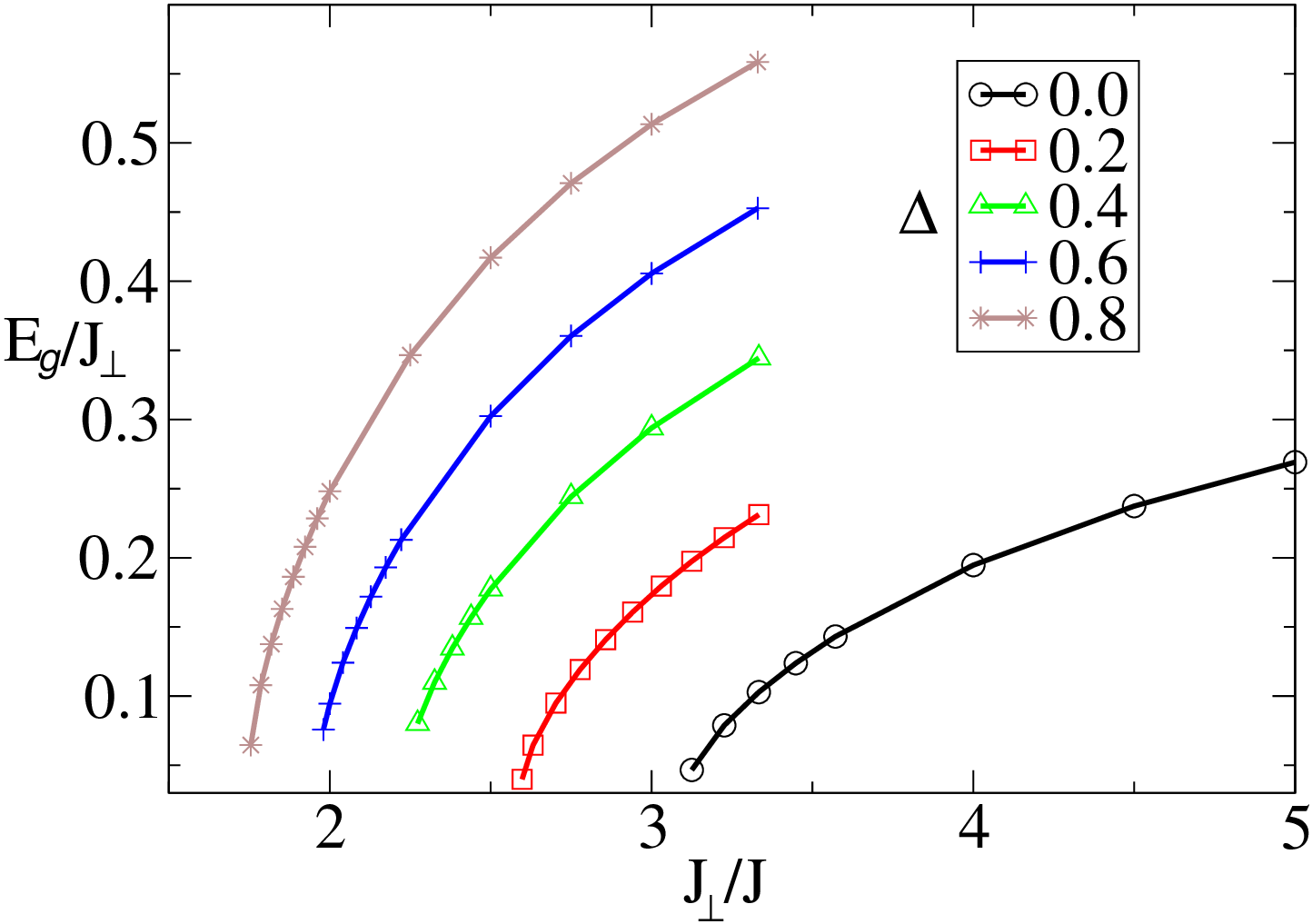}
\caption{\label{fig4} The energy gap ($E_g/J_{\bot}$) versus control parameter 
($J_\perp/J$) in the two dimensional lattice for $\delta=0$ and different values $\Delta$.}
\end{figure}

In the XY-case, the z-component of single particle excitation has approximately a dispersionless value $\omega_{z}(k)=\omega_{0}$. For the other components of excitations ($\omega_{x}=\omega_{y}$) the
dispersion relation show  a minimum at the reciprocal vector at the corner of the BZ, 
e.g., $Q=(\pi,\pi)$. In Fig.~\ref{fig4} we have plotted the energy gap ($E_{g}/J_{\perp}$) versus control parameter $J_{\perp}/J$. For all values of $\Delta$ the gap vanishes at the critical point $(\frac{J_{\perp}}{J})_c$, where the transition from Kondo singlet to the antiferromagnetic phase occurs. We have presented
the numerical values of the critical point in Table \ref{t1} for $\delta=0$ and different values of
intersite anisotropy, $\Delta=0.0, 0.2, 0.4, 0.6, 0.8, 0.9, 1.0$. In this case the critical points $(J_\perp/J)_c$
have also been plotted in Fig.~\ref{fig6} versus $\Delta$ (solid line). In Table~\ref{t1}, 
we have also compared the critical point with the mean field results \cite{langari}.
The gap exponent for different anisotropies have also been presented in Table~\ref{t1} and Fig.~\ref{fig7}. Our results in this case show that the anisotropy in the local interaction ($\Delta$) does not change the qualitative behavior of the Kondo necklace model on the 2D lattice. This is concluded  from numerical values for the gap exponent $\phi$ which are independent of $\Delta$.

\subsection{General anisotropic case $\delta\neq0$}
\begin{figure}[ht]
\includegraphics[width=8cm]{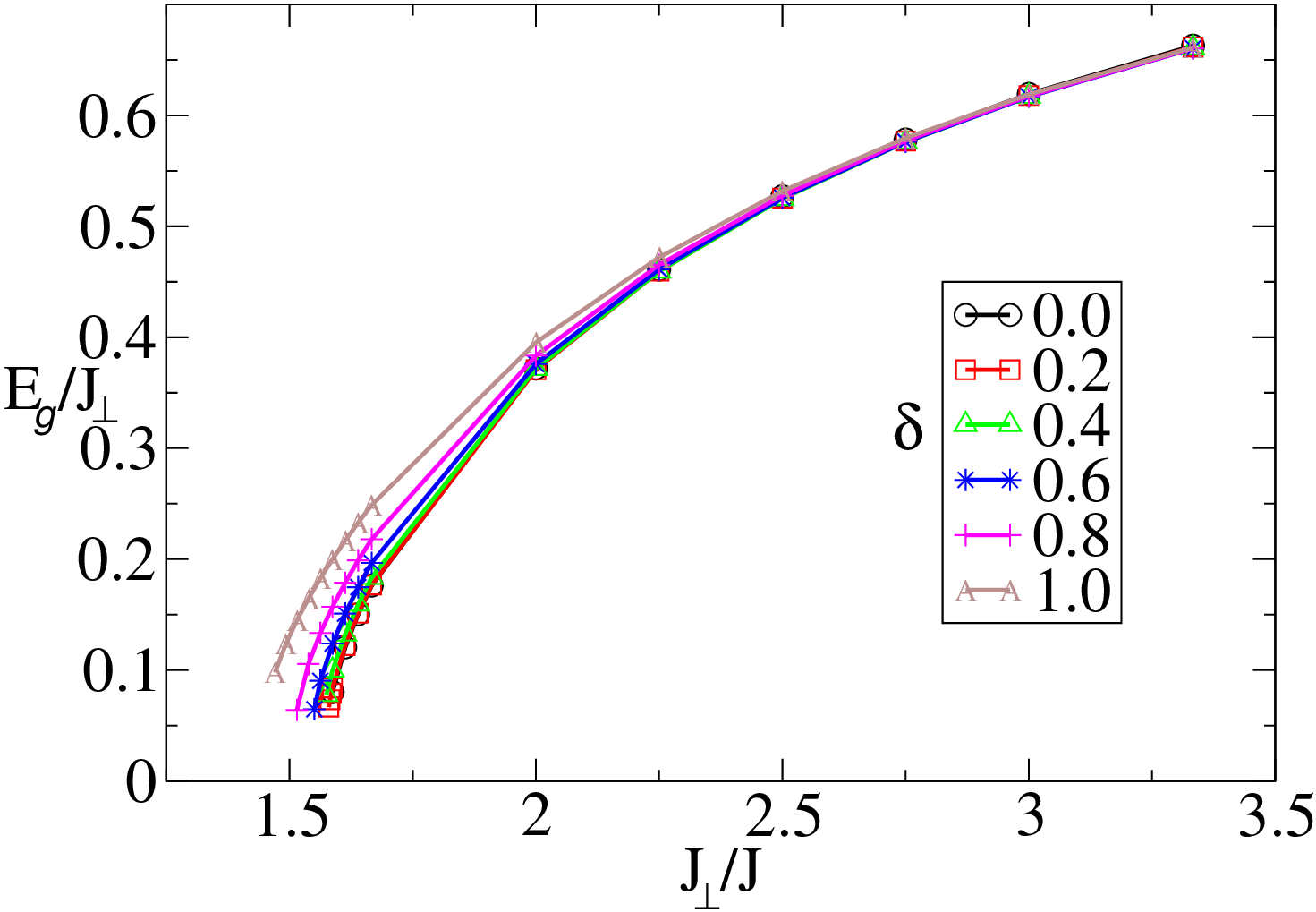}
\caption{\label{fig5} The energy gap ($E_g/J_{\bot}$) versus control parameter ($J_\perp/J$) in the 
two dimensional lattice for $\Delta=1$ and different values $\delta$.}
\end{figure}

\begin{figure}
\includegraphics[width=8cm]{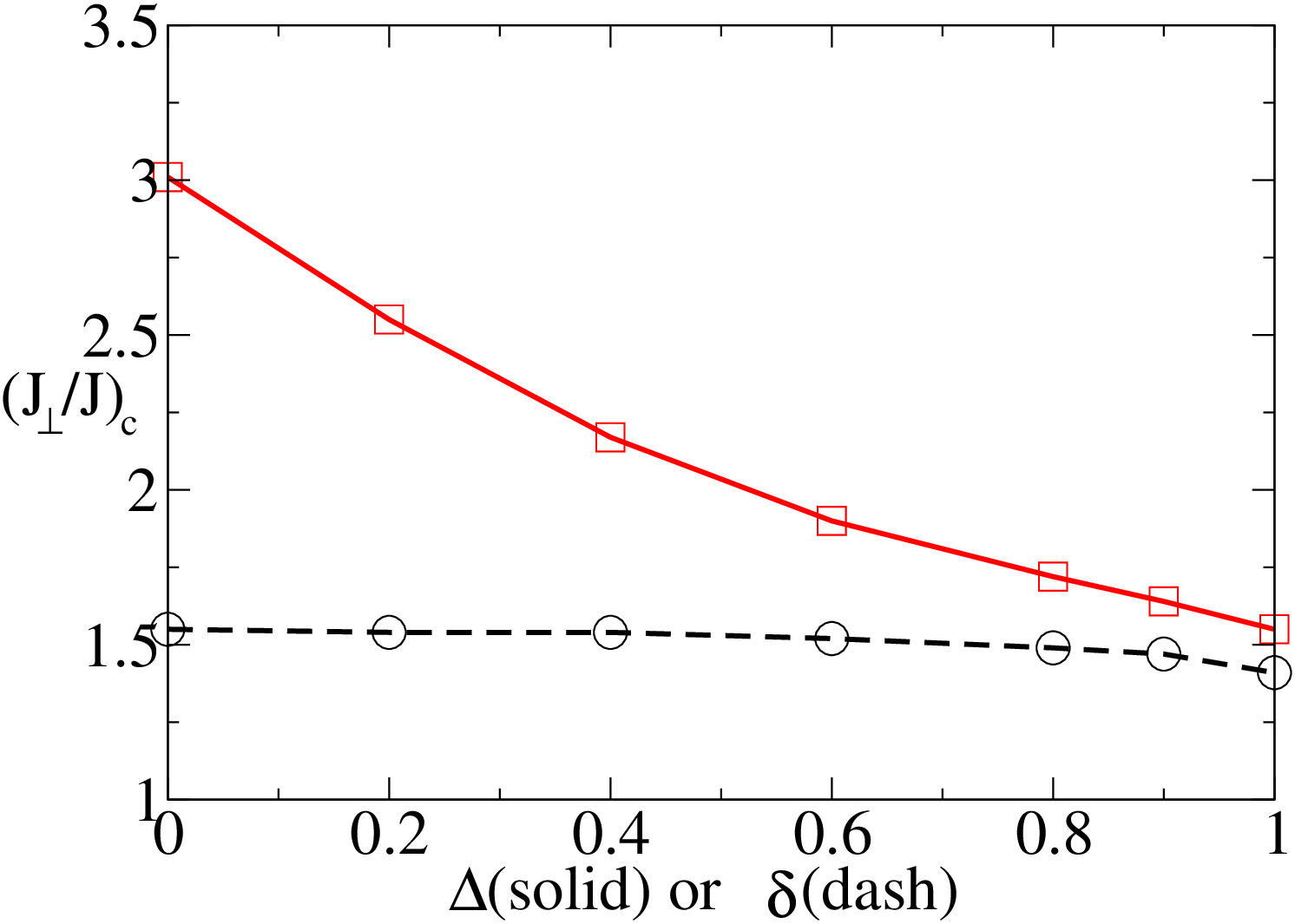}
\caption{\label{fig6}Dependence of the critical point $(J_\perp/J)_c$ on the local ($\Delta$, solid line) 
and itinerant ($\delta$, dashed line) anisotropies. For the solid line $\delta =0$ and for the dashed line $\Delta =1$.
The numerical values are given in Tables I and II.}
\vspace{0.5cm}
\end{figure}

\begin{figure}[ht]
\includegraphics[width=8cm]{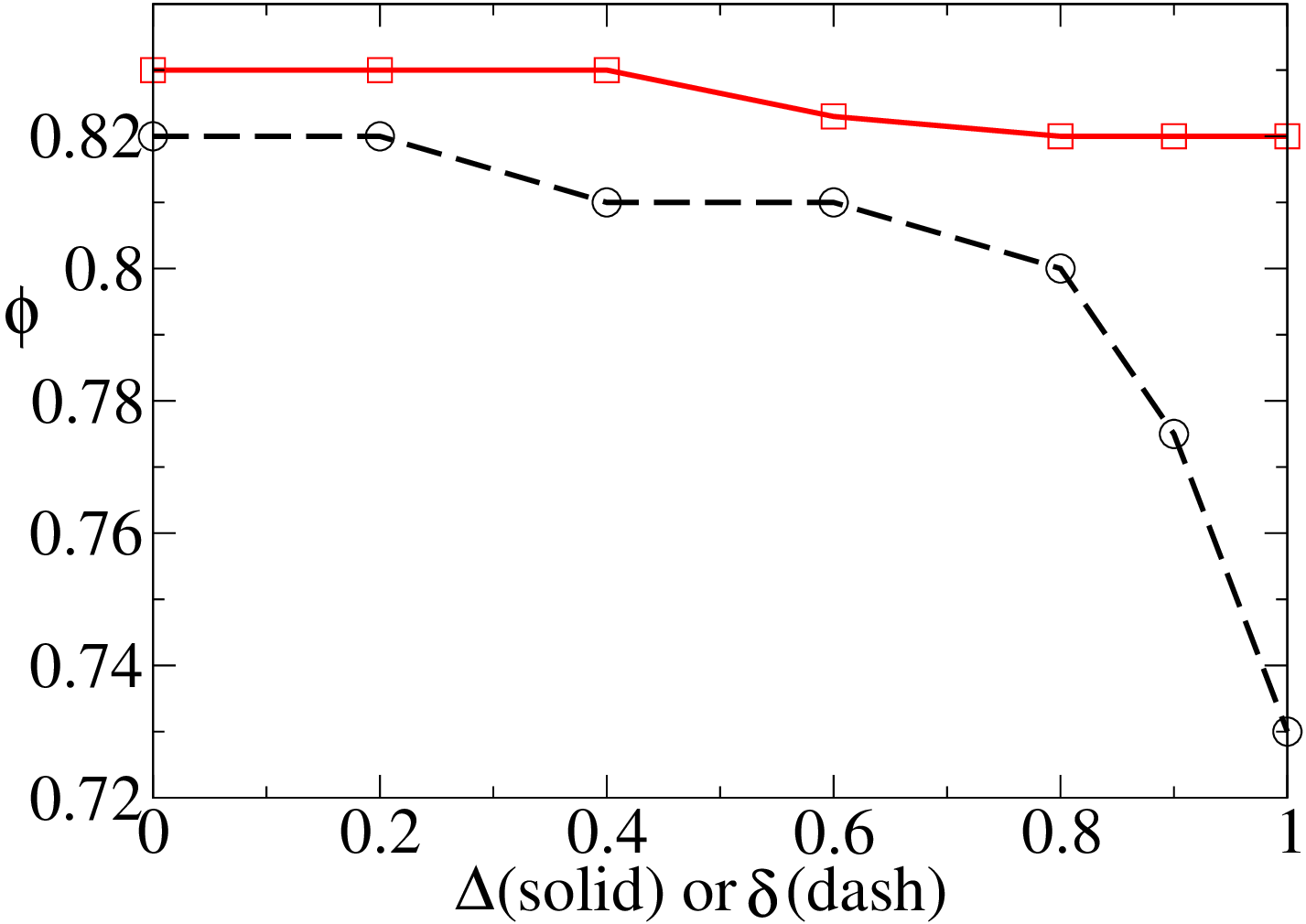}
\caption{\label{fig7}Dependence of the critical exponent ($\phi$) on the local 
($\Delta$, solid line) and 
itinerant ($\delta$, dashed line) anisotropies. 
For the solid line $\delta =0$ and for the dashed line $\Delta =1$. The numerical values are given in Tables I and II.}
\end{figure}

To study the effect of a nonzero anisotropy $\delta\neq0$ in the itinerant part we consider only the isotropic case of the local Kondo interactions, i.e., $J_{\perp x}=J_{\perp z}=J_{\perp},(\Delta=1)$. 
All of the three excitations show $k$-dependence in this case. The minimum excitation energy 
at Q wave vector defines the energy gap. For $0\leq\delta\leq1$, we have $E_{g}^{x}<E_{g}^{z}$ therefore we have plotted $E^x_{g}/J_{\perp}$ versus $J_{\bot}/J$ in Fig.~\ref{fig5}. In contrast to the case of
local anisotropy the effect of inter-site anisotropy on the critical point is weak and
$(J_{\bot}/J)_c$ changes only slightly. In Table~\ref{t2} and Fig.~\ref{fig6} we have shown
the calculated values of the critical point for different $\delta$. We have also presented the gap exponent for each 
anisotropy in Table~\ref{t2} and as a plot in Fig.~\ref{fig7}.
 We observe a rapid decrease of $\phi$ on approaching  the isotropic case ($\delta=1$). In
this case another soft mode ($\omega_{z}$) is added to the excitations at the critical
point and one needs a larger hopping strength ($J$) to reach the quantum critical point.
Moreover, the symmetry changes from U(1) for $0\leq \delta < 1$ to SU(2) at $\delta=1$.
Despite the weak dependence of the critical point on the anisotropy ($\delta$) the critical exponent of the gap changes with $\delta$ which is more pronounced at  $\delta=1$. However, the 
whole region of $0\leq \delta<1$ can be considered in a single universality class where
the gap exponent changes slightly while the change of exponent at $\delta=1$ 
signifies a different universality class by restoring the full spin rotational symmetry.

\begin{center}
\begin{table}[ht]
\caption{\label{t1} The critical point ($(\frac{J_{\bot}}{J})_c$) at which the singlet gap vanishes
for different values of intersite anisotropy ($\Delta$) and $\delta=0$. The second row shows the result from
Green's function approach and the third row gives the mean field values for the critical point \cite{langari}.
The gap exponent in the fourth row is obtained from the numerical evaluation of Eq.(\ref{e1060}).
The accuracy of data is $\pm 0.005$.
}\vspace{0.3cm} 
\begin{ruledtabular}
\begin{tabular}{cccccccc}
$\Delta$ &0.0&0.2&0.4&0.6&0.8&0.9&1.0\\
\hline
$(\frac{J_{\bot}}{J})_c$ (Green's function)&3.01&2.55&2.17&1.90&1.72&1.64&1.55\\
$(\frac{J_{\bot}}{J})_c$ (mean field)&2.86&2.38&2.04&1.78&1.59&1.51&1.43\\
$\phi$ (gap exponent)&0.83&0.83&0.83&0.82&0.82&0.82&0.82\\
\end{tabular}
\end{ruledtabular}
\end{table}
\end{center}

 \begin{center}
\begin{table}[ht]
\caption{\label{t2} The critical point ($(\frac{J_{\bot}}{J})_c$) at which the singlet gap vanishes
for different values of intersite anisotropy ($\delta$) and $\Delta=1$. The second row shows the result from
Green's function approach and the third row is the mean field values for the critical point.
The gap exponent in the fourth row comes from the numerical evaluation of Eqs.(\ref{e1060},\ref{e1070}).
The accuracy of data is $\pm 0.005$.
}\vspace{0.3cm} 
\begin{ruledtabular}
\begin{tabular}{cccccccc}
$\delta$ &0.0&0.2&0.4&0.6&0.8&0.9&1.0\\
\hline
$(\frac{J_{\bot}}{J})_c$ (Green's function)&1.55&1.54&1.54&1.52&1.49&1.47&1.41\\
$(\frac{J_{\bot}}{J})_c$ (mean field)&1.43&1.43&1.41&1.39&1.32&1.30&1.16\\
$\phi$ (gap exponent)&0.82&0.82&0.81&0.81&0.80&0.78&0.73\\
\end{tabular}
\end{ruledtabular}
\end{table}
\end{center}

\section{Discussion and Conclusion}
\label{sect:disc}

In this work we have carried the analysis of quantum critical behavior of the 2D anisotropic Kondo necklace model beyond the previous mean field treatment. The constraint on the bosonic excitations has been implemented with the help of a hard core boson term at every site instead of applying a global constraint by introducing a chemical potential in the mean field approach 
\cite{langari,thalmeier}.\\ 

The comparison of quantum critical points $(J_\perp/J)_c$ ($\delta =1$)as function of $\Delta$ in Table I shows that deviations
of the two methods are quite small, up to 10\% at the most. They are somewhat larger for the complementary case ($\Delta =1$) as function of $\delta$ (up to $\sim$ 17 \%), especially when the isotropic case $\delta =1$ is approached. The real difference and the advantage of the Green's function method appears when considering the critical exponent $\phi$ of the excitation gap $E_g$. In the mean field treatment the exponent is always  $\phi =1$ independent of the anisotropies. (Figs. 2 and 6 in Ref.~\onlinecite{langari}). On the other hand the present Green's function approach clearly leads to nontrivial exponents $\phi < 1$ as may already be seen by the direct comparison of Figs.~\ref{fig4},\ref{fig5} with those of Ref.~\onlinecite{langari} mentioned above. The calculated critical exponents $\phi$ are listed in Tables I and II and shown in Fig.~\ref{fig7}. Generally they lie around $\phi\simeq 0.80-0.83$. The most remarkable feature is the rapid reduction of $\phi$ when $\delta$ approaches the isotropic point $\delta =1$  where the universality class of the model changes from U(1)-xy to SU(2)-Heisenberg type. At this isotropic point we have $\phi\simeq 0.73$. Techniqually this means that  three instead of two soft modes at the AF wave vector $q_0$ appear which causes a rapid change in the gap exponent as described by Eqs.~(\ref{e1060},\ref{e1070}).\\

The  critical value   $(J_\perp/J)_c$ and exponent $\phi$ for the special isotropic case $(\delta,\Delta)=(1,1)$ has already been given in Ref.~\onlinecite{kotov}. From a numerical fit to the gap $E_g$ (Eq.~\ref{e551})  $(J_\perp/J)_c=1.39$ and $\phi=0.71$ was obtained which agrees reasonably well with our values   $(J_\perp/J)_c=1.41$ and $\phi=0.73$. In Ref.~\onlinecite{kotov} the isotropic case was also investigated for the true bilayer Hamiltonian. If we denote the inter-site coupling in the layer of localised spins (${\bf S}_i$) by $\lambda_S$ then $\lambda_S=0$ corresponds to the present KNM and $\lambda_S=J$ to the bilayer Hamiltonian. It was shown that the critical exponent $\phi$ for the isotropic case does not depend on $\lambda_S$. Applied to our anisotropic case we may conjecture that the critical exponents $\phi$ given in Tables I and II will also be valid for the anisotropic bilayer model  $\lambda_S=J$ although we have not done this calculation.\\

The quantum Monte-Carlo results \cite{sandvik} give $\nu=0.71$, where $\nu$ is the critical exponent 
of the correlation length close to quantum critical point of the 2D incomplete 
bilayer Heisenberg model (which is exactly the isotropic Kondo-necklace model). 
The dynamical exponent ($z$) relates the correlation length exponent ($\nu$) to the 
gap exponent ($\phi$) by $\phi=z \nu$. According to our results $\phi=0.73$ which is very close to the result 
for $\nu$ obtained in Ref.~\onlinecite{sandvik}, we conclude that the dynamical
exponent of the 2D isotropic Kondo-necklace model is $z=1$. As a consequence the effective dimension
of the model at the QCP is $D_{eff}=D+z=3$. This means that at the QCP the isotropic 2D KN model  corresponds 
to the universality class of the 3D classical Heisenberg model. This was also suggested in Ref.~\onlinecite{sandvik}.\\

It is also worthwile to discuss the different treatment of the bosonic constraint
$s_i^{\dag} s_i+ \sum_{\alpha}t_{i\alpha}^{\dag} t_{i\alpha} = 1$ (at every site i)
in the two methods in more detail since it is the source of the difference in the position of the quantum critical point  $(J_\perp/J)_c$ mentioned above. Their essential difference has already been explained in Sect.~\ref{sect:model}. In the mean field approach a Lagrange term with a chemical potential $\mu$ incorporating the constraint is added to the quadratic Hamiltonian $H_2$. Then the singlet and triplet boson operators
in the constraint are replaced by average amplitudes $\bar{s}$ and $\bar{t}_{\alpha}$ where the latter is identical to zero in the nonmagnetic phase. After diagonalization of $H_2$ the singlet amplitude $\bar{s}$ and the chemical potential $\mu$ are determined selfconsistently as function of $J_\perp/J$ by minimizing the total energy \cite{langari,thalmeier}. This means that $\mu$ and $\bar{s}$ will be slowly varying functions of the control parameter $J_\perp/J$. 
On the other hand the present Green's function approach corresponds to fixing these two parameters to  $(J_\perp/J)$ - independent constants  given by $\bar{s}=1$ and  $\mu=(J_\perp/4)(2+\Delta)$ which correspond to the mean field values for 
$J_\perp/J = 0$. Therefore on the level of H$_2$ there is no constraint implemented in the present approach. It rather appears through adding the $H_U$-term and computing its effect on the bosonic Greens functions in the limit $U \rightarrow\infty$. One might speculate whether it would be an improvement to start from the self-consistent mean field solution for $\bar{s}$ and only then impose the further constraint with the  $H_U$-term . However this cannot be justified easily. Once the constraints has been used in mean field level the fluctuations of the mean field singlet and triplet amplitudes should rather be unconstrained.\\

However, the hard core repulsion on bosonic excitations which is imposed by $H_U$, ensures that only
one triplet can be excited on each bond. It justifies the dominant contribution to the excitation 
spectrum of the model. While, in the mean field approach there is no restriction to have more
than one triplet excitation on each bond and the constraints
$s_i^{\dag} s_i+ \sum_{\alpha}t_{i\alpha}^{\dag} t_{i\alpha} = 1$ should be statisfied only when averaged over all bonds.
In other words, the Green's function approach  may be expected to give an improved excitation spectrum. This is
the reason why the critical exponent of the excitation gap obtained by Green's function approach is more accurate.

The calculation of critical gap exponents presented here presents a contribution to understanding the quantum critical behavior of the anisotropic Kondo necklace model. An extension to the antiferromagnetic side of the quantum critical point would be desirable to allow for a full comparison with the mean field results in Ref.~\onlinecite{langari}. However this will demand an even larger technical effort than in the present work.



\appendix\section*{Appendix A}\setcounter{section}{2}
The triplet scattering vertex in Eq.(\ref{e130}) is determined by the following equation
\begin{equation} 
\Gamma_{\alpha\beta,\alpha\beta}(\overrightarrow{K},\omega)=\frac{U}{1-iU
(2\pi)^{-4}\int
d^{3}\overrightarrow{Q}dQ_{0}G_{\alpha\alpha}^{0}(Q)G_{\beta\beta}^{0}(K-Q)\Gamma_{\alpha\beta,\alpha\beta}(\overrightarrow{K},\omega)}.
\label{e160}
\end{equation}
By substituting the normal noninteracting Green function Eq.(\ref{e100}) in the integral in Eq.~(\ref{e160}) we have
\begin{eqnarray}
\int dQ_{0}G_{\alpha\alpha}^{0}(Q)G_{\beta\beta}^{0}(K-Q)= \nonumber \\
\int
dQ_{0}(\frac{u_{\overrightarrow{Q},\alpha}^{2}}{Q_{0}-\omega_{\overrightarrow{Q},\alpha}+i\delta}
-\frac{v_{\overrightarrow{Q},\alpha}^{2}}{Q_{0}
+\omega_{\overrightarrow{Q},\alpha}-i \delta}) \times \nonumber \\
(\frac{u_{\overrightarrow{K}-\overrightarrow{Q},\beta}^{2}}{K_{0}-Q_{0}-\omega_{\overrightarrow{K}-\overrightarrow{Q},\beta}+i\delta}
-\frac{v_{\overrightarrow{K}-\overrightarrow{Q},\beta}^{2}}{K_{0}-Q_{0}+\omega_{\overrightarrow{K}-\overrightarrow{Q},\beta}-i\delta}).
\label{e150}
\end{eqnarray}
By using Cauchy's formula and extending the above integration over
the complex plane $Q_{0}$ we compute the residues of the simple poles
of the integrand. We found that the terms proportional to 
$u_{\overrightarrow{K}-\overrightarrow{Q},\beta}^{2}v_{\overrightarrow{Q},\alpha}^{2}$and
$v_{\overrightarrow{K}-\overrightarrow{Q},\beta}^{2}u_{\overrightarrow{Q},\alpha}^{2}$
give zero contribution because one pole is above the upper half plane and the other pole is below the lower half plane. After taking the limit $U\longrightarrow\infty$ then we obtain Eq.(\ref{e170}) for the scattering matrix.


\appendix\section*{Appendix B}\setcounter{section}{3}
We have the following equations for x,z component of the normal and anomalous self-energies
\begin{eqnarray}
 \Sigma_{n,x}^{3}(k,0)&=&\frac{J^{2}}{2N}\sum_{q}\Big(\frac{1}{\omega_{q,x}+\omega_{k+q,z}}[u_{q,x}v_{q,x}(u_{k+q,z}^{2}+v_{k+q,z}^{2})(\xi_{k}\xi_{q}-2\delta\xi_{k}\xi_{k+q}-\xi_{q}^{2}+2\delta\xi_{q}\xi_{k+q})\nonumber\\
&&+(u_{q,x}^{2}u_{k+q,z}^{2}+v_{q,x}^{2}v_{k+q,z}^{2})(-\delta^{2}\xi_{k+q}^{2}+\delta\xi_{q}\xi_{k+q})\nonumber\\
&&+(v_{k+q,z}^{2}u_{q,x}^{2}+u_{k+q,z}^{2}v_{q,x}^{2})(-\xi_{k}^{2}+2\xi_{k}\xi_{q}-\xi_{q}^{2})\nonumber\\
&&+u_{k+q,z}v_{k+q,z}(u_{q,x}^{2}+v_{q,x}^{2})(\delta\xi_{k}\xi_{k+q}-\delta^{2}\xi_{k+q}^{2})\nonumber\\
&&+2u_{k+q,z}v_{k+q,z}u_{q,x}v_{q,x}(\xi_{k}^{2}-\delta\xi_{k}\xi_{k+q}-\xi_{k}\xi_{q}+\delta\xi_{q}\xi_{k+q})]\nonumber\\
&&+\frac{1}{\omega_{q,z}+\omega_{k+q,x}}[u^{z}_{q}v^{z}_{q}(u_{k+q,x}^{2}+v_{k+q,x}^{2})(\delta\xi_{k}\xi_{q}-2\xi_{k}\xi_{k+q}-\delta^{2}\xi_{q}^{2}+2\delta\xi^{q}\xi^{k+q})\nonumber\\
&&+(u_{q,z}^{2}u_{k+q,z}^{2}+v_{q,z}^{2}v_{k+q,z}^{2})(-\xi_{k+q}^{2}+\delta\xi_{q}\xi_{k+q})\nonumber\\
&&+u_{k+q,x}v_{k+q,x}(u_{q,z}^{2}+v_{q,z}^{2})(\xi_{k}\xi_{k+q}-\xi_{k+q}^{2})\nonumber\\
&&+2u_{k+q,x}v_{k+q,x}u_{q,z}v_{q,z}(\xi_{k}^{2}-\delta\xi_{k}\xi_{k+q}-\xi_{k}\xi_{q}+\delta\xi_{q}\xi_{k+q})\nonumber\\
&&+(v_{k+q,x}^{2}u_{q,z}^{2}+u_{k+q,x}^{2}v_{q,z}^{2})(-\xi_{k}^{2}+2\xi_{k}\xi_{q}-\xi_{q}^{2})]\Big).
\label{e280}
\end{eqnarray}
 For the anomalous x component self-energy we obtain
\begin{eqnarray}
 \Sigma_{a,x}^{3}(k,0)&=&\frac{J^{2}}{2N}\sum_{q}\Big(\frac{1}{\omega_{q,x}+\omega_{k+q,z}}[u_{q,x}v_{q,x}(u_{k+q,z}^{2}+v_{k+q,z}^{2})(-\delta\xi_{k}\xi_{k+q}+\delta^{2}\xi_{k+q}^{2})\nonumber\\
&&+u_{k+q,z}v_{k+q,z}(u_{q,x}^{2}+v_{q,x}^{2})(2\delta\xi_{k}\xi_{k+q}+\xi_{q}^{2}-\xi_{k}\xi_{q}-2\delta\xi_{q}\xi_{k+q})\nonumber\\
&&+(v_{k+q,z}^{2}u_{q,x}^{2}+u_{k+q,z}^{2}v_{q,x}^{2})(-\xi_{k}^{2}+\delta\xi_{k}\xi_{k+q}-\delta\xi_{q}\xi_{k+q}+\xi_{k}\xi_{q})\nonumber\\
&&+2u_{k+q,z}v_{k+q,z}u_{q,x}v_{q,x}(\delta^{2}\xi_{k+q}^{2}-\delta\xi_{q}\xi_{k+q}+\xi_{k}^{2}+\xi_{q}^{2}-2\xi_{q}\xi_{k})]\nonumber\\
&&+\frac{1}{\omega_{q,z}+\omega_{k+q,x}}[u_{q,z}v_{q,z}(u_{k+q,x}^{2}+v_{k+q,x}^{2})(-\xi_{k}\xi_{k+q}+\xi_{k+q}^{2})\nonumber\\
&&+u_{k+q,x}v_{k+q,x}(u_{q,z}^{2}+v_{q,z}^{2})(2\xi_{k}\xi_{k+q}+\delta^{2}\xi_{q}^{2}-\delta\xi_{k}\xi_{q}-2\delta\xi_{q}\xi_{k+q})\nonumber\\
&&+2u_{k+q,x}v_{k+q,x}u_{q,z}v_{q,z}(\xi_{k+q}^{2}-\delta\xi_{q}\xi_{k+q}+\xi_{k}^{2}+\delta^{2}\xi_{q}^{2}-2\delta\xi_{q}\xi_{k})\nonumber\\
&&+(v_{k+q,x}^{2}u_{q,z}^{2}+u_{k+q,x}^{2}v_{q,z}^{2})(-\xi_{k}^{2}+\xi_{k}\xi_{k+q}-\delta\xi_{q}\xi_{k+q}+\delta\xi_{k}\xi_{q})]\Big).
\label{e290}
\end{eqnarray}
Similarly we can write the following expression for z component of the normal self-energy
 \begin{eqnarray} 
 \Sigma_{n,z}^{3}(k,0)&=&\frac{J^{2}}{N}\sum_{q}\frac{1}{\omega_{q,x}+\omega_{k+q,x}}[u_{q,x}v_{q,x}(u_{k+q,x}^{2}+v_{k+q,x}^{2})(\delta\xi_{k}\xi_{q}-2\delta\xi_{k}\xi_{k+q}-\xi_{q}^{2}+2\xi_{q}\xi_{k+q})\nonumber\\
&&+(u_{q,x}^{2}u_{k+q,x}^{2}+v_{q,x}^{2}v_{k+q,x}^{2})(-\xi_{k+q}^{2}+\xi_{q}\xi_{k+q})\nonumber\\
&&+(v_{k+q,x}^{2}u_{q,x}^{2}+u_{k+q,x}^{2}v_{q,x}^{2})(-\delta^{2}\xi_{k}^{2}+2\delta\xi_{k}\xi_{q}-\xi_{q}^{2})\nonumber\\
&&+u_{k+q,x}v_{k+q,x}(u_{q,x}^{2}+v_{q,x}^{2})(\delta\xi_{k}\xi_{k+q}-\delta^{2}\xi_{k+q}^{2})\nonumber\\
&&+2u_{k+q,x}v_{k+q,x}u_{q,x}v_{q,x}(\delta^{2}\xi_{k}^{2}-\delta\xi_{k}\xi_{k+q}-\delta\xi_{k}\xi_{q}+\xi_{q}\xi_{k+q})].
\label{e300}
 \end{eqnarray}
 For the corresponding anomalous self-energy we will have
\begin{eqnarray}
 \Sigma_{a,z}^{3}(k,0)&=&\frac{J^{2}}{N}\sum_{q}\frac{1}{\omega_{q,x}+\omega_{k+q,x}}[u_{q,x}v_{q,x}(u_{k+q,x}^{2}+v_{k+q,x}^{2})(-\delta\xi_{k}\xi_{k+q}+\xi_{k+q}^{2})\nonumber\\
&&+u_{k+q,x}v_{k+q,x}(u_{q,x}^{2}+v_{q,x}^{2})(2\delta\xi_{k}\xi_{k+q}+\xi_{q}^{2}-\delta\xi_{k}\xi_{q}-2\xi_{q}\xi_{k+q})\nonumber\\
&&+(v_{k+q,x}^{2}u_{q,x}^{2}+u_{k+q,x}^{2}v_{q,x}^{2})(-\delta^{2}\xi_{k}^{2}+\delta\xi_{k}\xi_{k+q}-\xi_{q}\xi_{k+q}+\delta\xi_{k}\xi_{q})\nonumber\\
&&+2u_{k+q,x}v_{k+q,x}u_{q,x}v_{q,x}(\xi_{k+q}^{2}-\xi_{q}\xi_{k+q}+\delta^{2}\xi_{k}^{2}+\xi_{q}^{2}-2\delta\xi_{q}\xi_{k})].
\label{e310}
\end{eqnarray}

\newpage

\newpage

\end{document}